\documentclass[12pt]{article}

\usepackage{setspace}   
\usepackage[labelformat=empty]{caption} 
\usepackage{fullpage}
\usepackage{xcolor}
\usepackage{graphicx}
\usepackage{mathtools}
\usepackage{authblk}
\usepackage{soul}  

\graphicspath{{Imagens/}}

\title{\textbf{Topological valley transport at the curved boundary of a folded bilayer graphene}}

\author{E. Mania$^{1,2}$}
\author{A. R. Cadore$^{1}$}
\author{T. Taniguchi$^3$}
\author{K. Watanabe$^3$}
\author{L. C. Campos$^{1,*}$}
\affil{$^1$Physics Department, Federal University of Minas Gerais, 30123-970, Brazil \\
	$^2$Physics Department, State University of Feira de Santana, 44036-900, Brazil\\
	$^3$National Institute for Materials Science, Namiki, 305-0044, Japan\\
	
	$^*$Correspondence and material requests should be addressed to L.C. Campos: lccampos@fisica.ufmg.br}

\date{}

\begin{document}

\maketitle
\doublespacing

\par This is a pre-print of an article published in Communications Physics. The final authenticated version is available online at: https://doi.org/10.1038/s42005-018-0106-4

\begin{abstract}
	\textbf{The development of valleytronics demands long-range electronic transport with preserved valley index, a degree of freedom similar to electron spin. A promising structure for this end is a topological one-dimensional (1D) channel formed in bilayer graphene (BLG) under special electrostatic conditions or specific stacking configuration, called domain wall (DW)\cite{Martin2008,Yao2009,Jung2011,Zhang2013,Vaezi2013}. In these 1D channels, the valley-index defines the propagation direction of the charge carriers and the chiral edge states (kink states) are robust over many kinds of disorder\cite{Jung2011}. However, the fabrication of DWs is challenging, requiring the design of complex multi-gate structures\cite{Li2016,Lee2017} or have been producing on rough substrates\cite{Ju2015,Jiang2018}, showing a limited mean free path. Here, we report on a high-quality DW formed at the curved boundary of folded bilayer graphene (folded-BLG). At such 1D conducting channel we measured a two-terminal resistance close to the quantum resistance $R = e^2/4h$ at zero magnetic field, a signature of kink states. Our experiments reveal a long-range ballistic transport regime that occurs only at the DW of the folded-BLG, while the other regions behave like semiconductors with tunable band gap.} 
\end{abstract}


Electric charge and spin are intrinsic quantum properties of electrons and so far, are the basis of electronics. Likewise, charge carriers in two-dimensional (2D) hexagonal crystals have an additional electronic degree of freedom, the valley-index, associated to degenerate bands at the inequivalent K and K' points in the Brillouin Zone (BZ). The valleytronic field proposes the creation of a new class of dissipationless electronic devices based on the manipulation of the valley-indices like valley filters and valley valves\cite{Martin2008,Qiao2011,Lee2012,Pan2015}. One interesting 2D crystal with useful features for valleytronics is BLG. The material is a tunable semiconductor and contains low lattice defects that prevents inter-valley scattering. Moreover, it holds topological properties when its inversion symmetry is broken by the application of transverse electric field. In this condition, a topological invariant is defined, the integer index called Chern number, with important implications on the quantum properties of BLG. For instance, it gives rise to the observation of the valley Hall Effect in graphene\cite{Gorbachev2014,Sui2015,Shimazaki2015}, which is a topological phase where gapless edge states labelled by opposite valley-indices counter-propagate at the boundaries of the insulating bulk\cite{Xiao2007}. One important aspect of the Chern number in BLG is that its sign depends either on the valley-index as well as on the sign of the band gap (interlayer energy difference), which can be changed by inverting the electric field direction or by inverting the stacking order of the material\cite{Xiao2007,Martin2008,Yao2009,Jung2011,Vaezi2013}. Such control of the band gap of BLG allows the design of topological 1D interfaces between regions with opposite Chern numbers - a domain wall - where strongly confined edge states called kink states are predicted\cite{Martin2008,Yao2009,Jung2011,Zhang2013,Vaezi2013}. The kink states have several useful features for valleytronics. There are two spin-degenerate kink states per valley and they are chiral, meaning that the propagation direction in the DW is defined by the valley-index. Such chiral edge states are robust for almost any kind of boundary configurations of the domains (except perfect armchair) and the topological protection inhibits backscattering from smooth disorder potentials\cite{Jung2011}. If valley-mixing is suppressed in the DW, such as by reducing the short-range disorder like edge defects and substrate corrugation, a dissipationless electrical conduction with conserved valley-index is expected.

\begin{figure}[h]
	\begin{center}
		\includegraphics{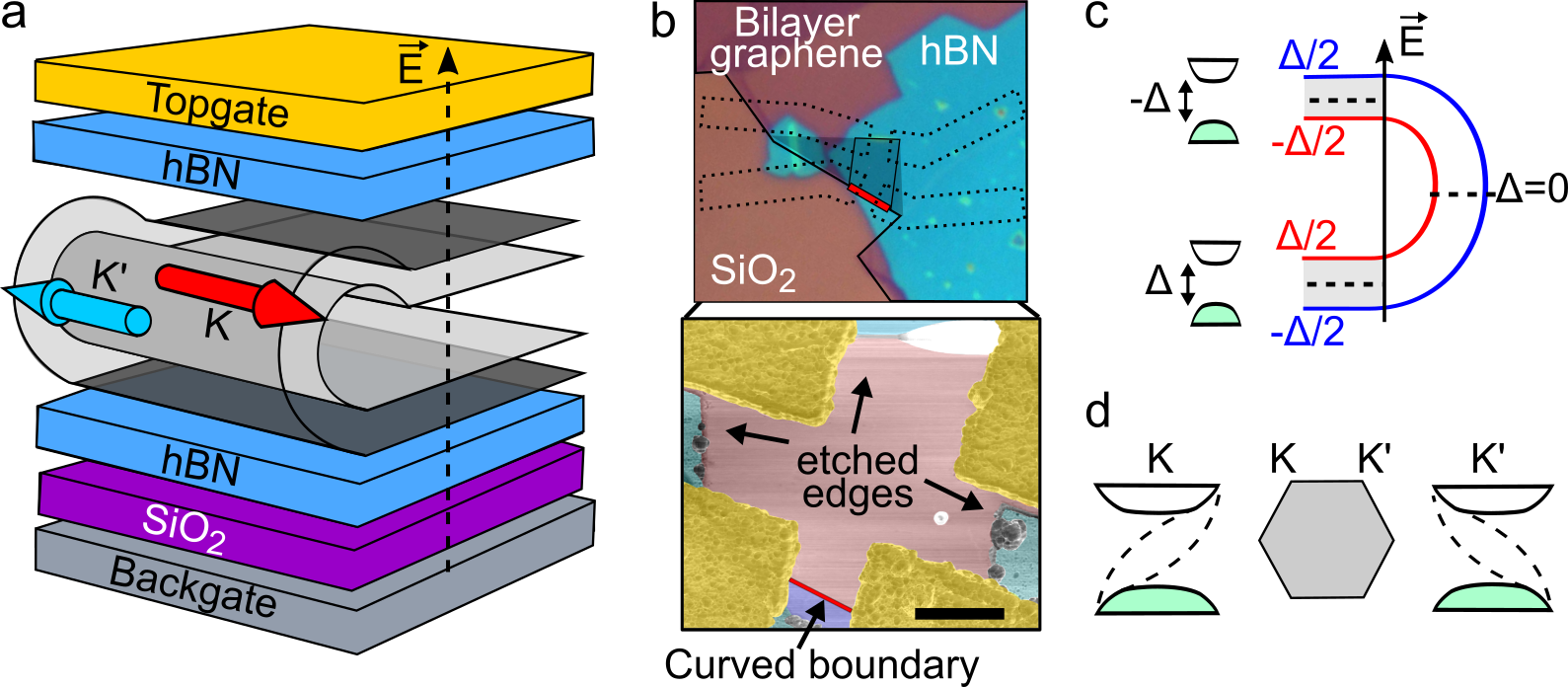}
		\caption{\textbf{Figure 1. Valleytronic device based on a folded bilayer graphene. a,} Components of the folded-BLG valleytronic device. The folded-BLG is sandwiched by hBN crystals that separate the material from the metallic gates. Under a transverse electric field, the bulk behaves likely semiconductor with band gap and a topological 1D conducting channel forms at the curved boundary, where the valley-index defines the direction of propagation. \textbf{b,} Optical image of a folded-BLG transferred on top of the bottom hBN flake and a false-colour AFM image of device 1 before the transference of the top hBN. Dashed lines indicate the position of the electric terminals. The AFM measurement reveals that the curved boundary is free from contamination of fabrication processes. Scale bar: 1 $\mu$m. \textbf{c,} Electrostatic potential energy ($\Delta/2$) of the BLG layers, calculated relative to the centre of each BLG. The layer energies reverse sign from the bottom BLG to the top BLG and vanish across the curved boundary, where the electric field is parallel to the layers. This variation of the electrostatic potential energy enable the formation of a domain wall at the curved boundary. \textbf{d,} Illustration of the pair of kink states localized at inequivalent points K and K' in the BZ of BLG, having opposite group-velocities for different valleys. In the domain wall, these edge states propagate in opposite one-dimensional directions due their chiral nature.}
		\label{fig:Figure1} 
	\end{center}
\end{figure}

To date, there are two routes to investigate kink states in BLG flakes. One exploits DWs formed along stacking faults (AB-BA boundaries). However, so far, such DWs have been only produced in BLG placed on top of rough SiO$_\text{2}$ substrates\cite{Ju2015,Jiang2018} showing limited mean free path. The other possible way is by designing complex gate-controlled topological channels, which requires a very precise alignment of the bottom and top gates\cite{Li2016,Lee2017}. Here, we observe strong evidence of kink states in high-quality DW formed at the curved boundary of a folded-BLG. Such compact geometry provides several advantages: the DW is atomically narrow, a variety of techniques enable the controlled production of such folded structures\cite{Li2006,Annett2016} and this architecture simplify the fabrication of valley-filters and valley-valves using fewer metallic gates. Moreover, we show that the topologically protected electronic transport is robust up to room temperature and shows a mean free path (MFP) up to the length of 20 $\mu$m at low temperatures, which is one of the longest MFP ever reported in a DW.

To introduce our valleytronic device, in the Fig. 1a we show a cartoon with some typical components of the device such as the folded-BLG, the metallic gates and the dielectrics. The folded-BLG is encapsulated in between two hexagonal boron nitride (hBN) crystals. The bottom hBN is placed on top of a SiO$_{\text{2}}$/Si$_{\text{++}}$ wafer, such that the Si$_{\text{++}}$ is a highly p-type doped silicon used as a backgate. The top hBN is covered by a metallic gate composed by Cr/Au. To illustrate our fabrication process, in the top image of Fig. 1b we show a typical heterostructure of a naturally folded-BLG on top of a flat hBN crystal, before patterning the metallic contacts (fabrication details are discussed in Methods and Supplementary Section 2). The dashed lines in this picture indicate the position of such electric terminals: two of them stand on the curved boundary and two of them are placed on the etched edges. The bottom image of Fig. 1b shows a false-colour AFM topography measurement of device 1. Here we present device 1 after the cleaning process and before encapsulating with a top hBN crystal. From this measurement, we see the high quality and cleanness of the device, which prevents short-range scattering along the 1D channel. These good conditions are provided either by the flat surface of the hBN crystals as well as by the efficacy of the mechanical cleaning method to remove contamination from the fabrication process (Method Section).

In the Fig. 1c we present a scheme that describes the electrostatic conditions imposed to the folded-BLG by application of gate potential. The transverse electric field breaks the inversion symmetry of each BLG (bottom sheet and top sheet of the folded-BLG), which leads to an energy band gap ($\Delta$)\cite{McCann2006} defined as the layer energy difference between the top graphene layer and the bottom graphene layer of BLG. At the bottom BLG sheet, the top graphene layer acquires a relative energy $+\Delta/2$, calculated relative to the centre of the sheet, and the bottom layer acquires a relative energy $-\Delta/2$. This electrostatic energy distribution inverts in the other BLG sheet on top. The bottom graphene layer (former top layer) now acquires a relative energy $-\Delta/2$ and the top layer (former bottom layer) acquires a relative energy $+\Delta/2$. Then, the band gap inverts its sign from the bottom to top BLG and, consequently, a valley at these different BLGs holds opposite Chern numbers. In this condition, the curved boundary of the folded-BLG transforms into a DW. We use the model proposed by Martin et al.\cite{Martin2008} to demonstrate the emergence of kink states in such topological folded structure (Supplementary Section 1). In the Fig. 1d we illustrate the pair of spin-degenerate kink states localized at the points K and K' of the BZ, having opposite group-velocities at different valleys. These chiral edge states propagate in the curved boundary along a 1D direction defined by the valley-index, as illustrated in Fig 1a. 

One of the main achievements of this work is the measurement of a quantization of the two-terminal resistance (R) along the curved boundary near of the quantum resistance $R = 6.45$ k$\Omega$ (or $R = e^2$/$4h$) at $B = 0$ T and $T = 1.2$ K. Such result is a remarkable evidence of kink states, since the conductance of the 1D channel is governed by a ballistic transport regime related to a pair of chiral edge states spin-degenerated\cite{Martin2008,Yao2009,Jung2011}. This result is presented in Fig. 2, which shows raw data of $R$ as a function of the backgate voltage ($V_{\text{BG}}$) and the topgate voltage ($V_{\text{TG}}$) measured at electric contacts placed along the etched edge (Fig. 2a) and along the curved boundary (Fig. 2b). Both measurements exhibit a diagonal line that shows a strong dependence of resistance with gate potentials. Along these diagonal lines, the electrostatic condition defined by the gate potentials set zero charge in the BLGs, called the charge neutrality point (CNP). The resistance at such diagonal lines are defined as follows: for the electric measurements realized on the etched edge it will be called $R^{\text{EE,CNP}}$ and for the electric measurements realized on the curved boundary it will be called $R^{\text{CB,CNP}}$. A better comparison of $R^{\text{EE,CNP}}$ and $R^{\text{CB,CNP}}$ is present in Fig. 2c, where we plot $R$ as a function of the displacement field $D$, obtained from the data showed in Fig. 2a and Fig. 2b. We converted the gate potentials to displacement field with the following formula: $D = (C_{\text{TG}}V_{\text{TG}}-C_{\text{BG}}V_{\text{BG}})/\epsilon_0$, where $C_{\text{TG}}$ and $C_{\text{BG}}$ are, respectively, the capacitances per unit of area and charge of the top capacitor and bottom capacitor, and $\epsilon_0$ is the vacuum permittivity. From data present in Fig. 2c we note that the monotonic increasing of $R^{\text{EE,CNP}}$ with $D$ reflects a tunable band gap caused by the broken inversion symmetry of BLGs\cite{Zhang2009,Taychatanapat2010}. In contrast, the $R^{\text{CB,CNP}}$ saturates near of the quantum resistance $R = h/4e^2$ for $|D|>1.6$ V/nm. This saturation of the resistance reveals that the DW formed at the curved boundary becomes electric isolated from the bulk of the folded-BLG and a ballistic transport regime governs the carrier motion at this 1D region. The quantization of resistance close to the quantum resistance show a robust valley transport, with backscattering strongly inhibited by the lack of short-range disorder along the channel. 

The strong suppression of backscattering in the DW formed along the curved boundary leads to a long MFP. We use the Landauer-B\"{u}ttiker formula\cite{datta1995} $R = R_{\text{0}}(1+L/L_{\text{MFP}})$ to calculate the MFP of the ballistic channel in our two devices. Here, $L_{\text{MFP}}$ is the MFP, $R_{\text{0}} = h/4e^2$ is the quantum resistance and $L$ is the length of the channels: $L = 1$ $\mu$m for device 1 and $L = 1.75$ $\mu$m for device 2 (Supplementary Section 4). Since the experiment shows a resistance close to $R = e^2$/$4h$, we neglected any other residual resistances in this channel. The calculated MFP of the channels are: $L_{\text{MFP}} \sim 20$ $\mu$m for device 1 and $L_{\text{MFP}} \sim 17$ $\mu$m for device 2. Such long MFP show that the DW formed at the folded-BLG is comparable to the best topological channels created by gate-confinement\cite{Li2016,Lee2017} and at least, two-orders higher than the MFP reported in a DW of BLG on SiO$_\text{2}$\cite{Ju2015}.

\begin{figure}[h]
	\begin{center}
		\includegraphics{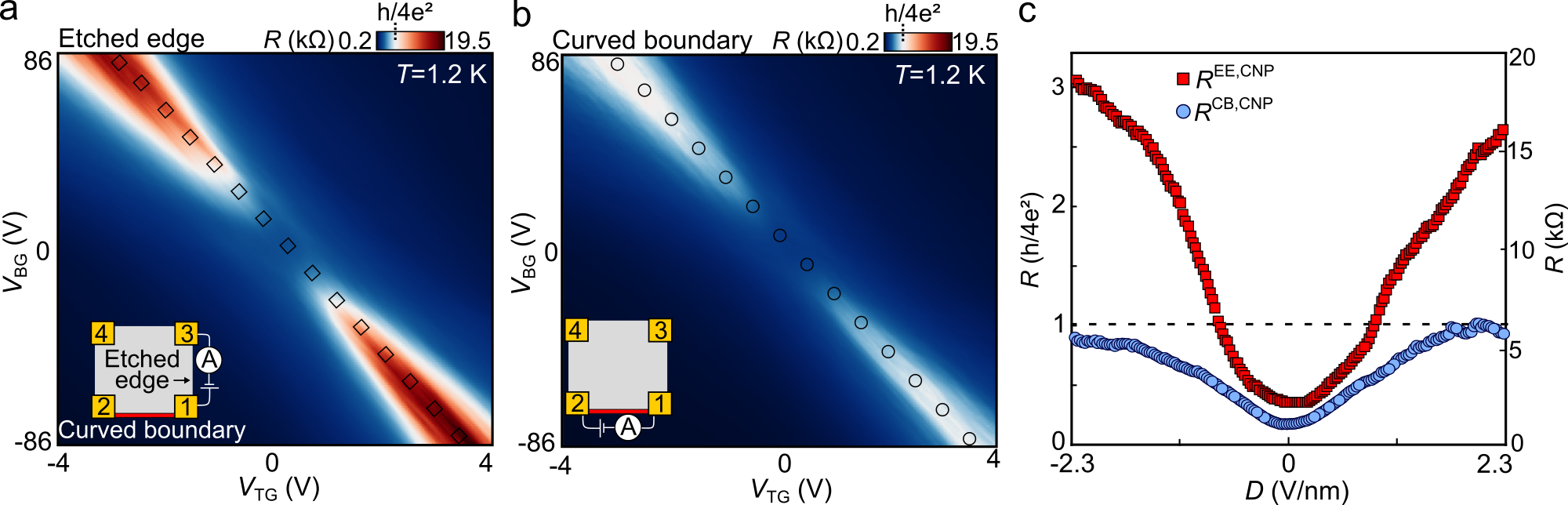}
		\caption{\textbf{Figure 2. Two-terminal electronic measurements: Evidence of a topological valley transport in the curved boundary of the folded-BLG. a,b,} Two-terminal electrical measurements of $R$ vs $V_{\text{TG}}$ vs $V_{\text{BG}}$ in the etched edge (contacts 1-3) and curved boundary (contacts 1-2), respectively, at $T = 1.2$ K and $B = 0$ T. The insets show each particular measurement configuration. \textbf{c,} $R$ as a function of $D$ from data showed in the Fig. 2a and Fig. 2b. The resistance at the etched edge, $R^{\text{EE,CNP}}$, monotonically increases with $D$ indicating a semiconducting regime with a band gap. For sufficient electric insulation of the bulk (achieved at high $D$) the resistance at the curved boundary, $R^{\text{CB,CNP}}$, saturate close to the quantum resistance $R = h/4e^2$, an evidence of kink states and the suppression of backscattering in this 1D conducting channel.}
		\label{fig:Figure1} 
	\end{center}
\end{figure}

Further details about the mechanisms of conduction in the folded-BLG are obtained by using a four-terminal electronic configuration. In the Fig. 3, we show the raw data of the longitudinal resistance ($R_{\text{xx}}$) as a function of the $V_{\text{BG}}$ and $V_{\text{TG}}$ measured at contacts placed along the etched edge (Fig. 3a) and the curved boundary (Fig. 3b). Again, both measurements show distinct diagonal lines where the resistance strongly depends on the gate potential. These dependence are better visualized in Fig. 3c, where we plot only the longitudinal resistances along the diagonal line as a function of $D$. Such resistances are called $R_{\text{xx}}^{\text{EE,CNP}}$ for the electric measurements realized at the etched edge and $R_{\text{xx}}^{\text{CB,CNP}}$ for the electric measurements performed at the curved boundary. We note that similarly to the two-terminal measurements, $R_{\text{xx}}^{\text{EE,CNP}}$ continuously increase with the displacement field. It changes from $R_{\text{xx}}^{\text{EE,CNP}}$ $\sim 1$ k$\Omega$ up to $R_{\text{xx}}^{\text{EE,CNP}}$ $\sim 12$ k$\Omega$, an expected behavior related to the tunable semiconducting nature of the BLGs by the action of the transverse electric field. On the other hand, we measure a different trend at the curved boundary channel. We observe that $R_{\text{xx}}^{\text{CB,CNP}}$ decreases with $D$, changing from $R_{\text{xx}}^{\text{CB,CNP}}$ $\sim 400$ $\Omega$ down to $R_{\text{xx}}^{\text{CB,CNP}}$ $\sim 60$ $\Omega$ for high $D$. Such small longitudinal resistance is a distinct feature of a ballistic transport regime in a channel with quasi-absence of backscattering. It provides another important evidence that a ballistic charge motion was achieved at the DW in the curved boundary.

Next, we investigate the effect of temperature on the electric conduction at the different regions of the folded-BLG. In Fig. 3d and Fig. 3e we show measurements of $R_{\text{xx}}$ as function of $V_{\text{TG}}$, with the backgate voltage fixed at $V_{\text{BG}} = -86$ V, while we changed the temperature of the system from $T = 1.2$ K up to room temperature ($T = 300$ K). At low temperatures, the electric conduction in the curved boundary is governed by a ballistic transport regime, revealed by the measurement of a small longitudinal resistance. At same electrostatic condition the BLGs at the etched edge reveal a semiconducting regime. As showed in Fig. 3d, the elevation of temperature in the system let to a decreasing of $R_{\text{xx}}$ measured along the etched edge. Clearly, it is the expected behavior of a semiconducting regime dominated by thermally activated processes\cite{Taychatanapat2010}. A different feature is observed at the curved boundary, as illustrated in Fig. 3e. The longitudinal resistance of such region increases when temperature goes up. Such behavior revels that temperature promotes valley-mixing by phonon scattering and enhances the scattering of edge states localized at the curved boundary channel to BLG states that may conduct due to charge inhomogeneity. These measurements on different temperatures give a third evidence that chiral edge states at the curved boundary let to a different electronic transport regime when compared to the other regions of the folded-BLG.

\begin{figure}[h]
	\begin{center}
		\includegraphics{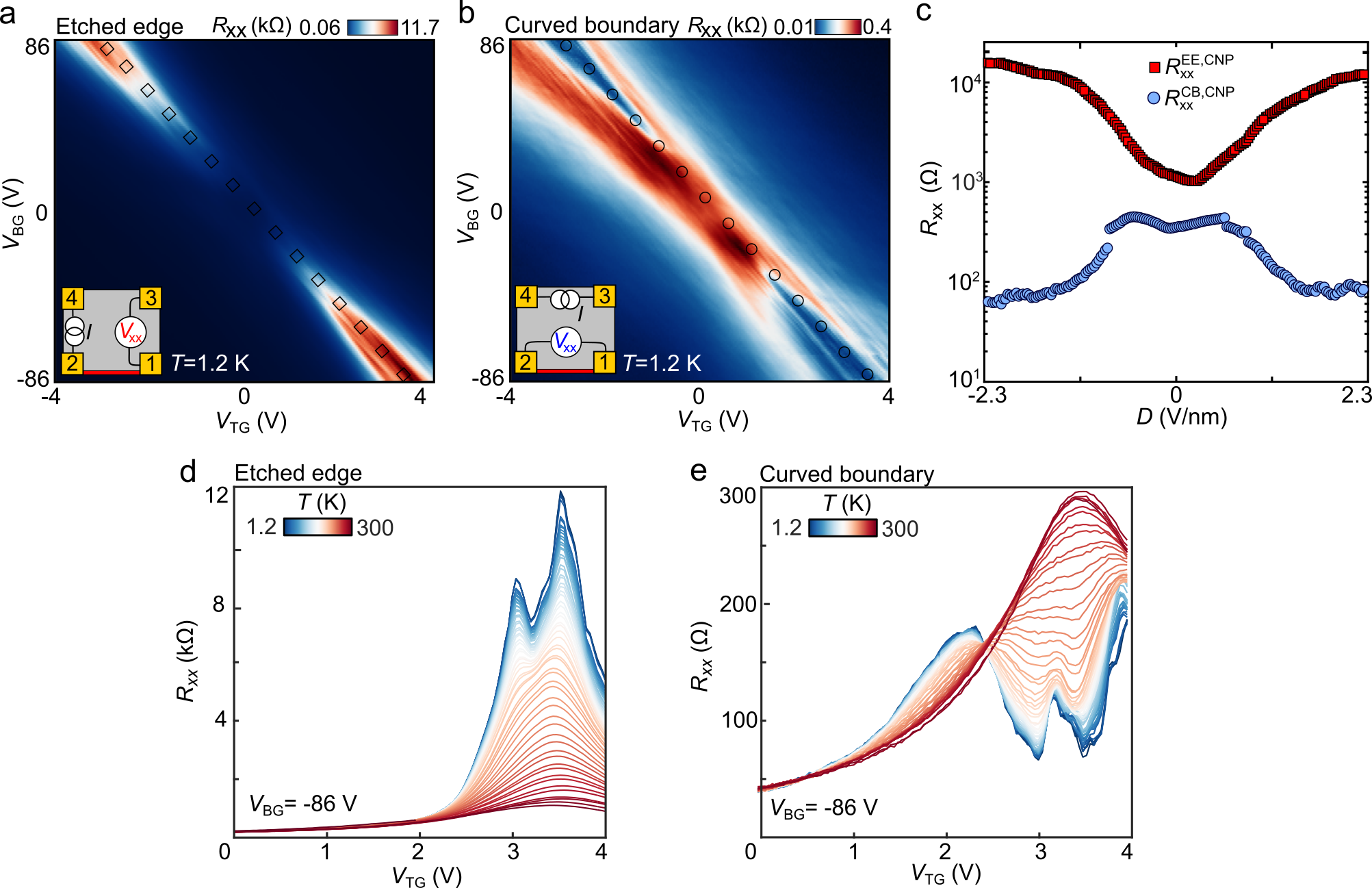}
		\caption{\textbf{Figure 3. Four-terminal electronic measurements and influence of temperature on the kink states.} \textbf{a,} and \textbf{b,} Four-terminal raw data of the longitudinal resistance $R_{\text{xx}}$ as a function of $V_{\text{TG}}$ and $V_{\text{BG}}$ in the etched edge and curved boundary, respectively. The insets show how the electronic measurements are performed. \textbf{c,} $R_{\text{xx}}$ as a function of $D$ when the BLGs are at the CNP. The resistances measured along the etched edge, $R_{\text{xx}}^{\text{EE,CNP}}$, are represented by the red squares and the resistances measured at the curved boundary, $R_{\text{xx}}^{\text{CB,CNP}}$, are represented by the blue circles. $R_{\text{xx}}^{\text{EE,CNP}}$ increases by one order of magnitude, while under same electrostatic conditions $R_{\text{xx}}^{\text{CB,CNP}}$ decreases down to a few tens of ohms, an expected behavior for the longitudinal resistance of a channel on the ballistic regime. \textbf{d,} and \textbf{e,} $R_{\text{xx}}$ in the etched edge and curved boundary, respectively, as a function of $V_{\text{TG}}$ for a fixed $V_{\text{BG}} = -86$ V, when the temperature of the system changes from $T = 1.2$ K up to $T = 300$ K. Such measurements reveals that different transport mechanisms governs the electric conduction at the etched edge region and at the curved boundary channel.}
		\label{fig:Figure1} 
	\end{center}
\end{figure}
\clearpage

In summary, our findings show the existence of topological chiral edge states in a domain wall formed at the curved boundary of a folded-BLG. We observe a strong suppression of valley scattering at this high-quality 1D channel that leads to a long-range ballistic conduction at zero-magnetic fields. Such novel platform contains elements to promote the development of dissipationless valleytronic devices and provides a new route to investigate graphene-based superconducting effects\cite{Schroer2015,Klinovaja2012} as well as Luttinger liquid interactions\cite{Killi2010}. 

\section*{Methods}

The heterostructures of folded-BLG sandwiched between hBN crystals are prepared with the following steps: we first employed the mechanical cleavage method to separate few layers of graphene from graphite flakes on top of a polymeric film of methyl methacrylate (MMA 495 C4). Next, we selected self-folded BLG samples and we transferred such flakes to top of clean hBN crystals supported on a 285-nm thick SiO$_{\text{2}}$/Si$_{\text{++}}$, where Si$_{\text{++}}$ is a highly doped Si wafer used as a metallic backgate. The fabrication of devices is divided into three main steps. First, we fabricated the electric terminals by using conventional electron beam lithography and thermal metalization of Cr/Au (1 nm/40 nm). We also used electron beam lithography and etching processes with oxygen plasma to define and shape our devices. Next, we used a mechanical cleaning method with an AFM probe\cite{Goossens2012} to remove any contamination of fabrication processes from the surface of the folded-BLG. We finished the fabrication by covering the device with another hBN flake and patterning a metallic top-gate. The electronic measurements are realized inside a cryogen system that enables the application of magnetic field up to $B = 7$ T. In our electronic measurements we normally operated at $T = 1.2$ K and we performed the measurements using a low-frequency ($f = 17$ Hz) Lock-in technique. In the two-terminal measurements we applied a constant bias ($V_{\text{bias}} = 1$ mV) between the contacts and we measured it electric current. The conductance is calculated by using the formula $G = I/V$. In the four-terminal measurements, a constant electric current ($I = 100$ nA) is applied between two electric terminals and a longitudinal voltage ($V_{\text{xx}}$) is measured in between the other electric terminals on the opposite side. The longitudinal resistance is calculated by using the Ohm's law $R_{\text{xx}}$=$V_{\text{xx}}/I$.

\section*{Acknowledgements}

The authors thank Javier D. Sanchez-Yamagishi, Hadar Steinberg and Marcos H. D. Guimaraes for the fruitful discussions and the paper revision. The authors acknowledge the support of LabNS and LabNano for the Raman and AFM measurements. E.M., A.R.C. and L.C.C. acknowledge the support of the Brazilian agencies: Conselho Nacional de Desenvolvimento Cientifico e Tecnologico (CNPq/MCTI), Coordenacao de Aperfeicoamento de Pessoal de Nivel Superior (Capes), Fundacao de Amparo a Pesquisa do Estado de Minas Gerais (FAPEMIG). K.W. and T.T. acknowledge support from the Elemental Strategy Initiative
conducted by the MEXT, Japan and the CREST (JPMJCR15F3), JST.

\section*{Author contributions}

L.C.C. conceived the main idea of the work. E.M. and A.R.C. fabricated the devices. K. W. and T.T. supplied the high-quality hBN flakes. E.M. and L.C.C. planned, realize the measurements, analyzed the data and wrote the manuscript. 

\section*{Competing Interests}

The authors declare no competing interests.

\bibliographystyle{naturemag}
\bibliography{references}

\end{document}


\maketitle
	\doublespacing

\par This is a pre-print of an article published in Communications Physics. The final authenticated version is available online at: https://doi.org/10.1038/s42005-018-0106-4

\section*{S1: Prediction of kink states at the curved boundary of a folded bilayer graphene}

\qquad This section describes the emergence of kink states at the curved boundary of a folded bilayer graphene (folded-BLG). We will show that the curved boundary is a topological interface equivalent to the one predicted by Martin at. al.\cite{Martin2008} in between two gapped bilayer graphene (BLG) regions under opposite transverse electric field. We first describe the low-energy effective Hamiltonian of the BLG when the electrostatic potential energy of its layers change in space (kink potential). Equation (1) shows the simplified two-component Hamiltonian\cite{McCann2006} acting in the space of wave functions related to non-dimer states $\Psi (x) =(\psi _{\text{A2}} (x),\psi _{\text{B1}} (x))$,

\begin{equation}\label{eq:teoria_hamilt_3}
	{H}_{\text{B}}=
	-\frac{1}{2m}
	\begin{pmatrix}
		0 & (\pi{^\dagger})^2\\
		\pi^2 & 0
	\end{pmatrix}
	+\xi
	\begin{pmatrix}
		\frac{1}{2}\Delta(x) & 0\\
		0 & -\frac{1}{2}\Delta(x)
	\end{pmatrix}\,,
\end{equation}

where $\xi$ is +1 for valley K and -1 for valley K'. The potential energy of each layer are $\epsilon _1$ and $\epsilon _2$, corresponding, respectively, to the bottom graphene layer and top graphene layer of the BLG sheet. We define the asymmetry between such on-site energies relative to the middle of the layers as a function of the interlayer energy difference: $\Delta=|\epsilon _1 -\epsilon _2|$, with $\epsilon _1=\Delta/2$ and $\epsilon _2=-\Delta/2$. When the electric field (E) flips up side down, as illustrated in the Fig. S1a, the layer energies change their sign. Fig. S1b shows the sketch of the energy changing across a kink region with an arbitrary width $d$. As described by Martin at. al.\cite{Martin2008}, a set of second-order differential equations is the result of applying the Hamiltonian of equation (1) in the wave functions, 

\begin{equation}\label{eq:teoria_hamilt_3}
	-u(x) \psi _{\text{A2}} + (\partial _{\text{x}} + p_{\text{y}})^2 \psi _{\text{B1}}=E \psi _{\text{A2}}\,,
\end{equation}

\begin{equation}\label{eq:teoria_hamilt_3}
u(x) \psi _{\text{B1}} + (\partial _{\text{x}} - p_{\text{y}})^2 \psi _{\text{A2}}=E \psi _{\text{B1}}\,,
\end{equation}

where we define $u(x)= m \Delta(x)$, the energy E is normalized ($E = 2 m E$) and we choose the particular valley K ($\xi =1$). The system has an analytical solution for the ideal kink profile ($d = 0$). Here we reproduce the numerical solution for the realistic profile ($d > 0$), using $u(x) = tanh(x/d)$ and $d = 1$. In the Fig S1c, we show that such solution is a pair of spin-degenerated states (kink states) at the valley K (another solution can be obtained for the valley K'). and the states of the gapped BLG (marked in green). 

\par Next, we show that a folded-BLG under a transverse electric field can be described by the Hamiltonian of equation (1). Fig. S1d shows the layer energies of each BLG for a particular direction of the electric field. Like in the single BLG case, the layer energies are represented as a function of the interlayer energy difference ($\Delta$) and calculated relative to the centre of each BLG. The layers energies of the bottom BLG sheet are similar to energies of the layers at the BLG on the left side of Fig. S1a. On the other hand, the folding reverse the position of non-dimer states in the top BLG sheet. Consequently, the layer energies reverse their sign, being compared to the BLG in the right side of Fig. S1a. In the curved boundary, the layers of BLG turns parallel to the electric field direction and $\Delta$ should vanish in the middle of this edge. Such a space-variation of $\Delta$ at the curved boundary as well as the change in the signal of the layer energies are similar to the change created by flipping the electric field up. Thus, we can describe the folded-BLG with the Hamiltonian present in equation (1) and we expect that kink states are confined in the DW formed at the curved boundary. 

\par To finish, let us discuss some important points. In our experiments we do not observe any evidence of the formation of superlattices and, therefore, we treat the folded bilayer graphene far from the curved boundary using a naive approximation of two independent sheets of BLG. As expected in our approximation, the BLGs behave like a semiconductor with tunable band gap. However, this band gap estimated from electrical measurements is smaller than reported in BLG\cite{Zhang2009,Taychatanapat2010}. Finally, in our naive approximation we do not consider any quantum effect that could be caused by the curvature of the curved boundary, which may be a subject of future theoretical studies.
 
\begin{figure}[h]
	\begin{center}
		\includegraphics{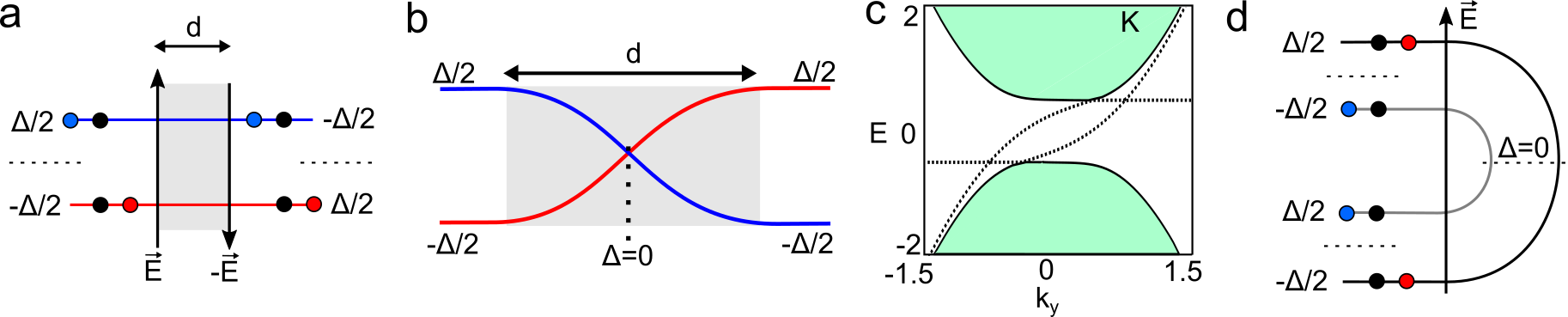}
		\caption*{\textbf{Figure S1. Model of a BLG and folded bilayer graphene under a kink potential.} \textbf{a,} A Bernal stacked BLG under opposite transverse electric fields. The electric field breaks the layer symmetry of the BLG, leading to different on-site layer energies (measured relative to the centre of the BLG). Such layer energies have opposite signals in each region due to the different electric field direction. \textbf{b,} Sketch of a kink potential region with arbitrary width $d$, showing the space variation of the layer energies. \textbf{c,} Numerical solution of equation (2) and equation (3) showing a pair of spin-degenerate states (kink states) around the valley K in the low-energy Brillouin Zone. The kink states have opposite group velocities in the valley K'. For comparison, in green we plot states of the conduction and the valence bands of the gapped BLG. \textbf{d,} A folded-BLG under a transverse electric field. The folding reverse the position of the non-dimer states from the bottom to the top BLG, inverting their layer energies. At the curved boundary, the interlayer energy $\Delta$ vanishes since the electric field is parallel to the layers. At such condition, the curved boundary is similar to the interface in between inverted electric fields in a single BLG.}
		\label{fig:figure01} 
	\end{center}
\end{figure}
\clearpage

\section*{S2: Fabrication steps}

\begin{figure}[h!]
	\begin{center}
		\includegraphics{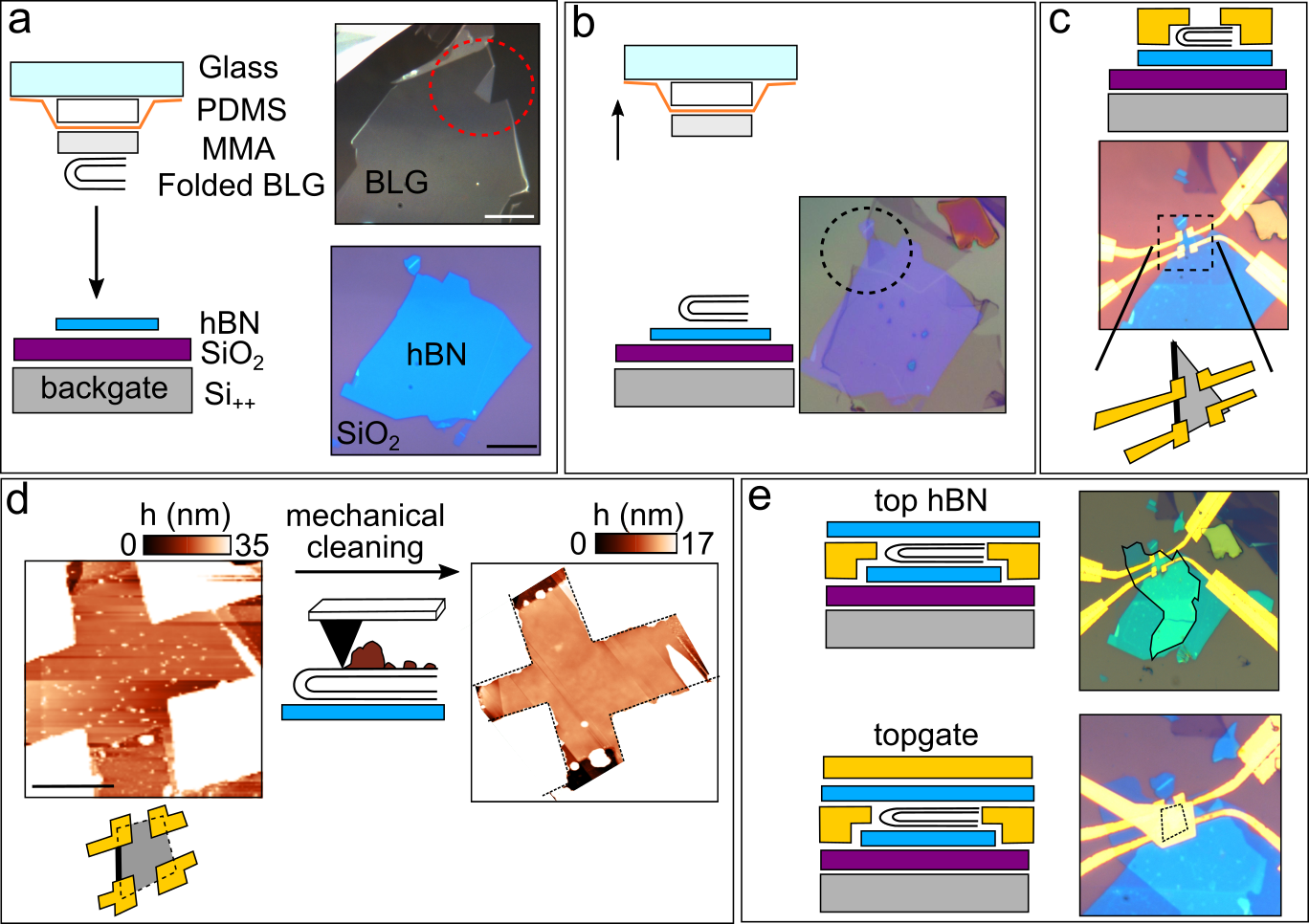}
		\caption{\textbf{Figure S2. Fabrication steps of a valleytronic device based on a folded bilayer graphene. a,} Left side: a schematic view of the transference process, with its main components: the membrane supporting the BLG flake and the substrates of device. Right side: Optical images of an exfoliated BLG flake supported on the membrane (the red dashed circle indicates a self-folded BLG) and the exfoliated hBN flake is shown on top of the SiO$_{\text{2}}$/Si$_{\text{++}}$ wafer, where  the Si$_{\text{++}}$ is a highly doped silicon. Scale bar: 10 $\mu\text{m}$. \textbf{b,} BLG was transferred on top of hBN with a dry transfer method. The optical picture shows the BLG as well its folded part transferred on top of the hBN flake. \textbf{c,} Metallic electric terminals were fabricated with conventional e-beam lithography and thermal metalization of Cr/Au (1nm/40nm). The optical picture and its scheme show the position of the electric terminals on the sample. \textbf{d,} Left panel: AFM topography picture after fabrication of the electric terminals. There are many contamination on top of the folded-BLG surface. Central panel: a schematic view of the mechanical cleaning method, which uses a AFM tip to remove the contamination from graphene surface\cite{Goossens2012}. Right panel: AFM topography picture after the cleaning procedure, showing that both folded-BLG surface as well its the curved boundary are free of contamination. \textbf{e,} Final device fabrication steps. Upper panel: shows a top hBN flake transferred on top of the sample. Bottom panel: shows a topgate fabricated in a way that cover both the folded-BLG and its curved boundary.}
		\label{fig:figure0} 
	\end{center}
\end{figure}
\clearpage

\section*{S3: Raman Characterization}

\begin{figure}[h]
	\begin{center}
		\includegraphics{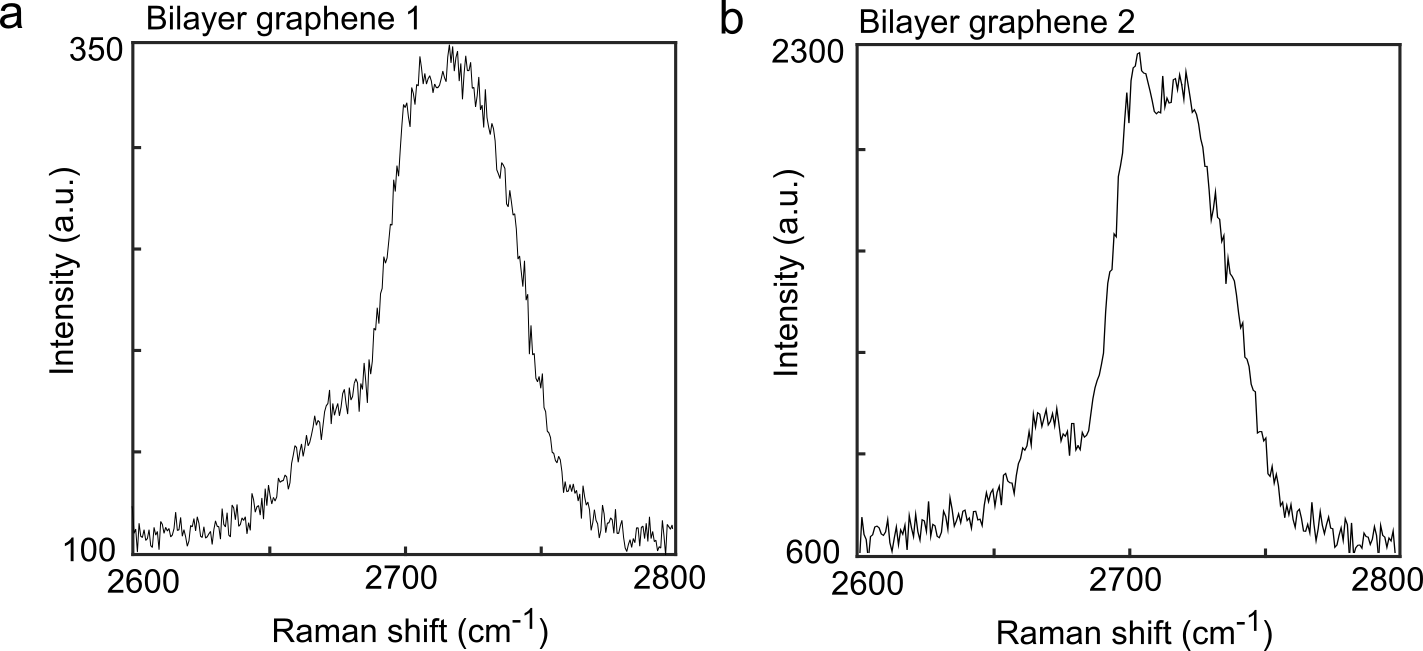}
		\caption{\textbf{Figure S3. Determination of the number of layers of graphene by using the Raman spectroscopy. a,} and \textbf{b,} show, respectively, the measured Raman 2D peak of the graphene flake of device 1 and device 2. The measurements are realized near of the folded region, using an excitation laser with the wavelength of 532 nm. For this wavelength, the shape of both measured 2D Raman peaks is quite similar to the expected shape of a BLG\cite{Malard2009}. It confirms that the folded region is composed by two BLG.}
		\label{fig:figure3} 
	\end{center}
\end{figure}
\clearpage

\section*{S4: Evidence of kink states in device 2}

\qquad We fabricate a second valleytronic device (device 2) and performed electronic measurements at the curved boundary of a folded-BLG. At this device we observed evidence of kink states, again revealed by the quantization of resistance close to the quantum resistance of $R = 6.45$ k$\Omega$ at zero-magnetic field. We present this device in Fig. S4a. The top image shows the optical picture of a folded-BLG flake transferred on top of a hBN crystal. The bottom image shows a AFM topography picture of device 2 after we have cleaned the folded-BLG surface. In Fig. S4b we present the raw data of two-terminal $R$ vs $V_{\text{BG}}$ vs $V_{\text{TG}}$ at $T = 1.2$ K measured from the electric contacts placed along the curved boundary. We performed two-terminal measurements only at the curved boundary because the contacts at the etched edges did not work properly. Like the measurements performed on device 1, the resistances at electrostatic conditions of the CNP are showed at the diagonal line ($R^{\text{CB,CNP}}$). In Fig. S4c we show these $R^{\text{CB,CNP}}$ vs $D$, obtained from the measurement present in the Fig. S4b. We note that $R^{\text{CB,CNP}}$ saturates close to the quantum resistance $R = h/4e^2$ for $|D| > 1.6$ V/nm, which indicates that a DW forms at the 1D curved boundary and the charge motion is governed by a ballistic transport regime. Using the Landauer-B\"{u}ttiker formula\cite{datta1995} $R = R_{\text{0}}(1+L/L_{\text{MFP}})$ we calculate the mean-free path of the channel: $L_{\text{MFP}} = 17.5$ $\mu$m. We next study the influence of a magnetic field $B$ on the electronic properties of the curved boundary channel. We chose the electrostatic condition ($D \sim 1.4$ V/nm) such that the resistance is saturated close to the ballistic resistance (see the arrow in Fig. S4c), and we ramped the magnetic field in the interval $0$ T$< B < 6.9$ T. Fig. S4d shows such resistance $R^{\text{CB,CNP}}$ as function of $B$. We note that $R^{\text{CB,CNP}}$ remains close to $R = e^2/4h$ for all $B$. At such conditions, the increase of the magneto resistance on the bulk lead to a suppression of backscattering, promoting the protection of the chiral edge states in the curved boundary.

\begin{figure}[h!]
	\begin{center}
		\includegraphics{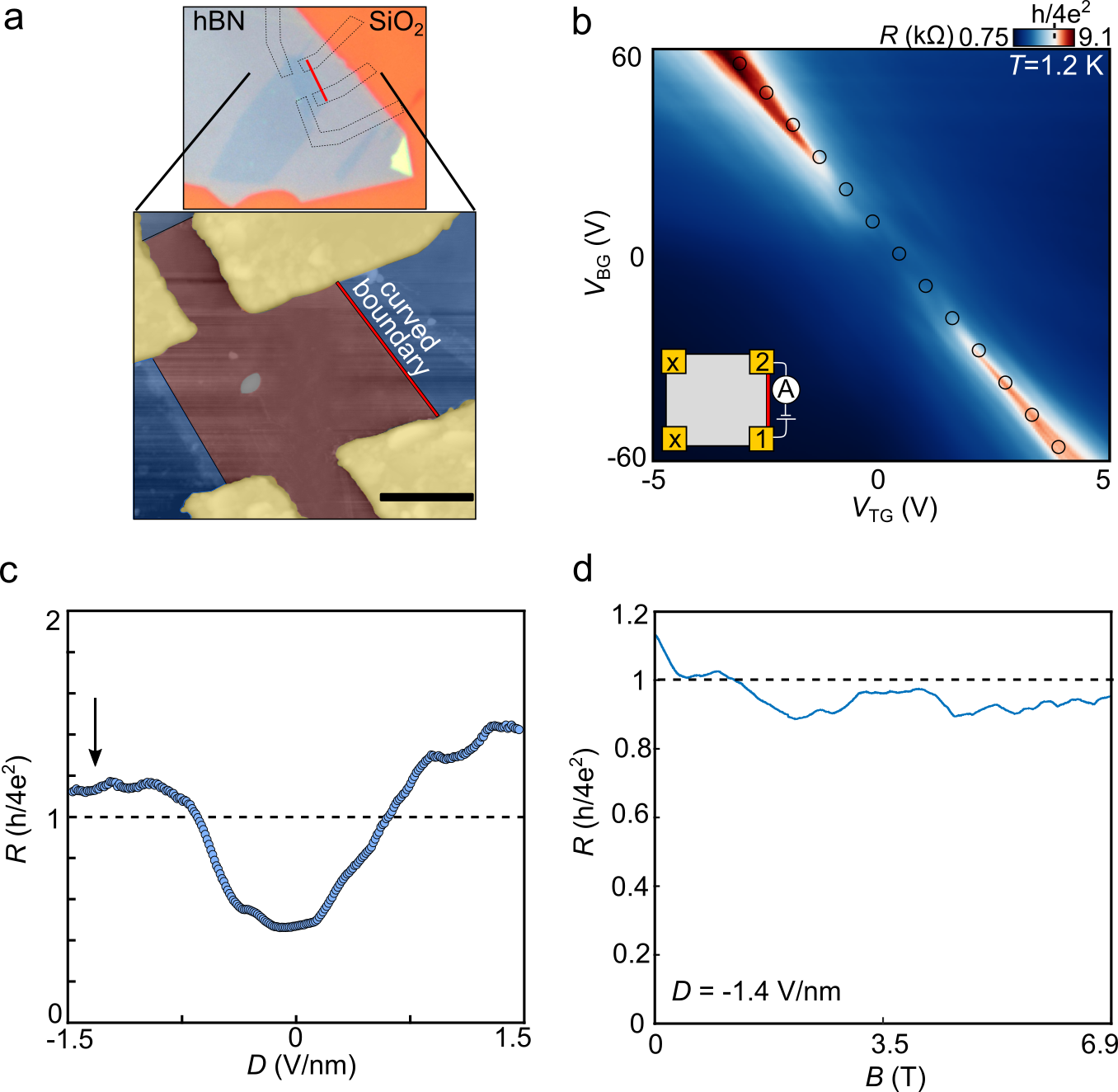}
		\caption{\textbf{Figure S4. Electronic transport properties of device 2. a,} Top image: Optical picture of a folded-BLG deposited on top of a hBN crystal. Bottom image: false-color AFM topographic measurement of device 2 after the cleaning process and before it was covered with a top boron nitride flake. The curved boundary and its surroundings are free from contamination of the fabrication process. Scale bar: 1 $\mu$m. \textbf{b,} Raw data of the two-terminal $R$ vs $V_{\text{BG}}$ vs $V_{\text{TG}}$ measured along the curved boundary at $T = 1.2$ K and $B = 0$ T. The inset show the electronic measurement configuration. \textbf{c,} $R^{\text{CNP}}$ vs $D$ of the data presented in Fig. S4b. The two-terminal resistance saturates near of the quantum resistance $R = e^2/4h$ for $D < -0.8$ V/nm. A similar behavior was observed in device 1 and is another important evidence of the kink states at the curved boundary of a folded-BLG. \textbf{d,} The $R^{\text{CNP}}$ at the saturation, indicated by the arrow of Fig. S4c, as function of B. For $B \sim 0.5$ T and $B \sim 3.5$ T the resistance of the channel approaches the quantum resistance, indicating that backscattering are inhibit in the channel.}
		\label{fig:figure4} 
	\end{center}
\end{figure}
\clearpage

\section*{S5: Influence of temperature on the kink states}

\begin{figure}[h]
	\begin{center}
		\includegraphics{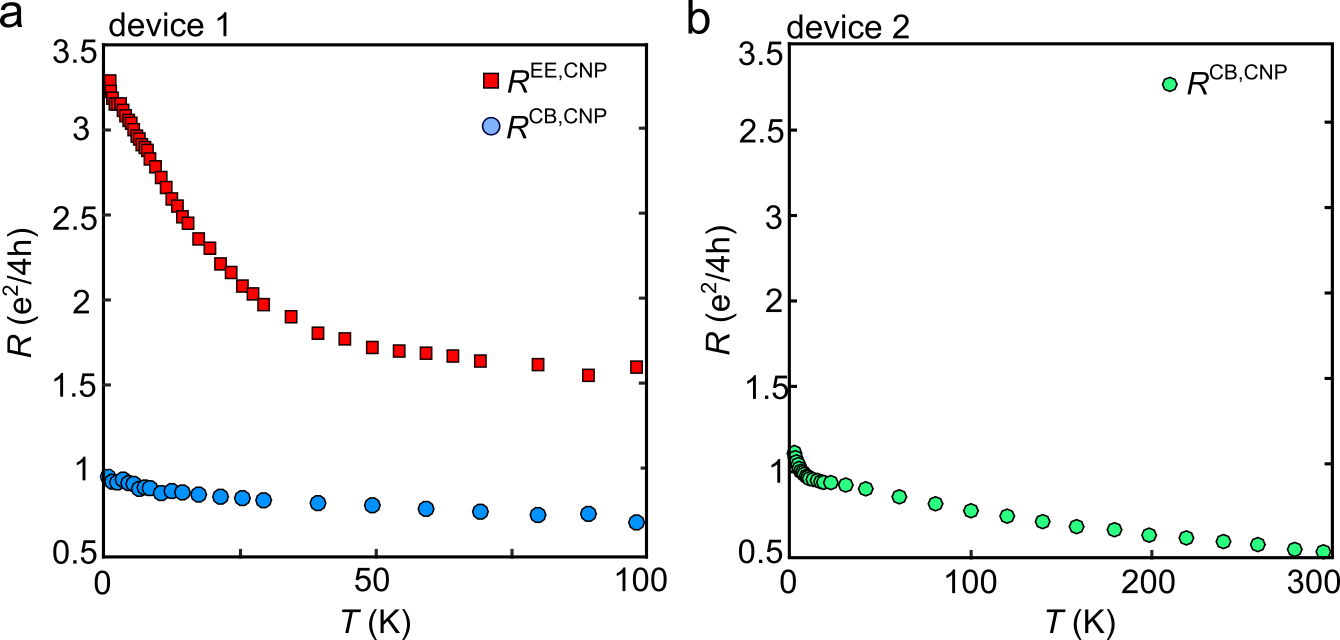}
		\caption{\textbf{Figure S5. Two-terminal electronic measurements as function of temperature. a,} Two-terminal resistance as function of temperature measured on device 1, along the etched edge (red square), $R^{\text{EE,CNP}}$, and curved boundary (blue circle), $R^{\text{CB,CNP}}$. \textbf{b,} Two-terminal resistance $R^{\text{CB,CNP}}$ measured in the curved boundary of device 2 for temperatures in the range of $1.2$ K$< T < 300$ K.}
		\label{fig:figure5} 
	\end{center}
\end{figure}

\qquad In this section, we describe the effects of temperature on the electronic transport properties of kink states at the domain wall in the curved boundary of the folded-BLG. In Fig. S5a we show the measurement of two-terminal resistance as function of temperature on device 1. We chose the electrostatic gate potentials such that we achieve the ballistic transport regime at the curved boundary. The blue circles are the resistances measured along the curved boundary, $R^{\text{CB,CNP}}$, while the red squares are the resistance measured along the etched edge at the electrostatic conditions of charge neutrality point, $R^{\text{EE,CNP}}$. The measurements are done for temperatures ramping from $T = 1.2$ K up to $T = 100$ K. We observe different mechanisms of conduction at the curved boundary and at the etched edge. The resistance $R^{\text{EE,CNP}}$ rapidly decreases when temperature goes up and show a small decreasing rate for higher temperatures. Such behavior is related to thermally activated conducting processes - typical for semiconducting materials. On the other hand, $R^{\text{CB,CNP}}$ slowly decreases with temperature, remaining near of the quantum resistance $R = e^2/4h$ even at $T = 100$ K. In the Fig. S5b, we show $R^{\text{CB,CNP}}$ measured along the electric contacts at the curved boundary of device 2. Here, we ramp the temperature of the system from $1.2$ K up to room temperature ($T = 300$ K). Again, we observe a slow decrease of $R^{\text{CB,CNP}}$ with temperature, which remains on the order of the quantum resistance even at room temperature. Such robust feature of topological valley transport at high temperatures is similar to the non-local signal of the valley Hall Effect\cite{Sui2015,Shimazaki2015}. We believe that a large band gap in the bulk of the folded-BLG will provide a better protection of the chiral edge states in DW, that may promote a stable operation of valleytronic devices at ambient conditions. 

\bibliographystyle{naturemag}
\bibliography{references}